\documentclass{article}

\usepackage{arxiv}

\usepackage[utf8]{inputenc} 
\usepackage[T1]{fontenc}    
\usepackage{hyperref}       
\usepackage{url}            
\usepackage{booktabs}       
\usepackage{amsfonts}       
\usepackage{nicefrac}       
\usepackage{microtype}      
\usepackage{lipsum}
\usepackage{graphicx}
\usepackage{subcaption}
\usepackage{mathtools}
\usepackage[export]{adjustbox}
\usepackage{algorithm}
\usepackage{algpseudocode}
\usepackage{xcolor}
\usepackage{listings}
\usepackage{mdframed}
\graphicspath{ {./images/} }
\usepackage{biblatex} 

\title{An evaluation of LLM code generation capabilities through graded exercises}

\author{
 Álvaro Barbero Jiménez \\
  Instituto de Ingeniería del Conocimiento and Universidad Autónoma de Madrid\\
  Madrid, Spain \\
  \texttt{alvaro.barbero@iic.uam.es}
}

\begin{document}
\maketitle

\begin{abstract}
Large Language Models have shown prominent capabilities in generating functional code from natural language descriptions. However, a standardized way to evaluate these capabilities in an objective and unbiased manner is still to be found. In this paper we review the current evaluation methods available to this end, and run a new evaluation of the performance of one state-of-the-art model (GPT4-o-mini) in solving curated coding challenges in 8 programming languages, obtained from Codewars, a software development community. Our analysis shows that the chance of success of the model has a positive correlation with the task difficulty, the popularity of the programming language being used and the time elapsed since the publication of the challenge. A further approximate explanatory analysis in terms of high-level features hints that while 46.6\% of the model performance could be attributed to task difficulty, a 37.4\% seems to be related to leakage of the challenge solutions into the model training set, while the remaining 16\% depends on the programming language. These results suggest that current evaluation methodologies might be overestimating the actual skill of Large Language Models for generating functional code.
\end{abstract}

\section{Introduction and Background}

\subsection{Large Language Models}

\textbf{Large Language Models} (LLMs) are deep neuronal networks built from Transformer blocks \cite{vaswani2023attentionneed}, which model the distribution of written text splitted into words pieces, also known as tokens. They are usually trained in an unsupervised way over massive amounts of text, generally obtained from internet crawls. Such training can be implemented by predicting tokens masked at random in a segment of text and configuring the Transformed layers to use bidirectional attention (encoder models) \cite{devlin2019bertpretrainingdeepbidirectional, liu2019robertarobustlyoptimizedbert, he2021debertadecodingenhancedbertdisentangled, distilbert}, or by issuing the model to predict the token following a given piece of text and configuring the attention masks in a left-to-right or causal way (decoder models) \cite{gpt-2, gpt-3, gpt-neo, bloom, gpt-4, llama, falcon, llama2, mistral, mixtral, qwen, stablelm2, llama3, gpt-4o, qwen2}. Some hybrid encoder-decoder architectures have also been proposed \cite{raffel2023exploringlimitstransferlearning}. While early LLMs mainly focused on the English language, recent models tend to be trained in a varied mixture of languages; however language-specific models also abound \cite{camembert, bertimbau, maria, rigoberta, albeto}.

The most valuable property of LLMs is that they allow for an effective \textbf{transfer learning} to a wide variety of language-related tasks, such as document classification, named entity recognition, summarization, question answering, information retrieval or natural language inference, to name a few. Such transfer can be performed by \textbf{fine-tuning} the model to a small supervised dataset related to the downstream task, generally producing better results than training a model from scratch over the same data.

Even more, it has been shown \cite{gpt-3} that pre-training a decoder LLM in a large enough dataset grants it some capabilities to perform well in downstream tasks without requiring a fine-tuning step, by means of \textbf{prompting} the model with a description in natural language of the task to perform (zero-shot inference), sometimes adding some input-output examples of the task to boost performance (few-shot inference). Since writing prompts requires no specialized knowledge in machine learning or the underlying details of the model, this method of interaction with LLMs has rapidly expanded to general users all over the world \cite{chatgpt, claude}.

However, for LLMs to be able to cater for the wide variety of natural language requests from all kinds of users, some specialized post-training steps have been shown to be necessary. \textbf{Instruction tuning} \cite{flan}, also termed Supervised Fine Tuning (SFT), improves the ability of the LLM to produce adequate responses to users, by fine-tuning it over datasets written by humans that show the desired outputs of the LLM for different user requests. While this method is effective, it is also expensive in terms of human labour. A more efficient way of collecting human inputs and using them for adapting the model is \textbf{Reinforcement Learning from Human Feedback} (RLHF) \cite{christiano2023deepreinforcementlearninghuman, stiennon2022learningsummarizehumanfeedback, ouyang2022traininglanguagemodelsfollow}, where the LLM generates two or more outputs for a given prompt, and a human rater ranks them in terms of quality; the data gathered this way is then used to update the LLM using reinforcement learning algorithms \cite{ppo, rafailov2024directpreferenceoptimizationlanguage}, so as to produce outputs more similar to those favoured by the humans.

Despite all these efforts, LLMs have also been shown to suffer from a number of failure modes and \textbf{lack of generalization outside their training data distribution}. LLMs, or more generally decoder Transformer models, make use of prompting as a way to align their predictions to a particular data mixture seen in pre-training, and thus fail to solve tasks not present in their pre-training or instruction-tuning data \cite{yadlowsky2023pretrainingdatamixturesenable,lu2024emergentabilitieslargelanguage}. 

\textbf{Biases towards generating responses more frequent in the training data} or in the few-shot prompts have also been observed \cite{zhao2021calibrateuseimprovingfewshot}. The position where the relevant information in the prompt is placed also turns out to be critical, as LLMs tend to have \textbf{primacy and recency biases} that favor using information at the beginning or the end of the prompt, an effect also known as "lost in the middle" \cite{liu2023lostmiddlelanguagemodels, zhao2021calibrateuseimprovingfewshot}. Even the format of the prompt itself can have significant impact in the generated text, leading to users creating all kinds of prompt variations in what has been termed "prompt engineering".

More generally, all models learned by gradient descent (which includes LLMs) have been shown to approximately behave as a kernel machine: they \textbf{implicitly memorize training data} and generate predictions as a linear combination of training examples, weighted by their similarity to the input being processed \cite{domingos2020modellearnedgradientdescent}. These evidences highlight the fact that, despite their resounding success, LLMs are restrained by their training data, much in the way of any other machine learning model.

\subsection{Evaluating LLMs}

Objectively \textbf{evaluating} the performance of LLMs poses a significant challenge, partly due to the wide range of tasks they can be applied to, but also because of their training procedure: internet crawls can contain copies of the very same datasets used for testing, thus overestimating their performance due to data leakage \cite{open-llm-leaderboard-v2}. 

A popular evaluation method is using a number of \textbf{evaluation datasets in the form of multiple choice tests}, in domains such as general knowledge, math or science. The LLMs are scored by checking if the correct choice for a question is the one for which the LLM assigns the highest generation probability \cite{open-llm-leaderboard-v2, eval-harness, zhou2023instructionfollowingevaluationlargelanguage, suzgun2022challengingbigbenchtaskschainofthought, hendrycks2021measuringmathematicalproblemsolving, rein2023gpqagraduatelevelgoogleproofqa, sprague2024musrtestinglimitschainofthought, wang2024mmluprorobustchallengingmultitask, open-llm-leaderboard-v1}; these benchmarks, however, need to be updated frequently to avoid leakage to new LLMs through updates in the crawls used in their training data.

Another approach involves using \textbf{crowd-sourced human evaluations} of model outputs in the form of preferences \cite{chiang2024chatbotarenaopenplatform}, ranking the models by the percentage of times they were preferred, or by an Elo rating. Although generally considered to be the most accurate method for measuring the capabilities of an LLM, they are expensive to carry out and requires of domain experts to implement evaluations in highly specialized fields.

A third approach employs \textbf{automated evaluation metrics} that measure the semantic similarity of the LLM response versus a gold standard, or assign some other kind of score to the response. These metrics need to make use of yet another LLM, either specialized in embedding texts into a latent space in which semantic similarity can be numerically measured \cite{sas}, or in generating a performance report of the adequateness of the response \cite{prometheus2}. While convenient and cheap, these kind of metrics have been shown to be brittle and suffer from biases \cite{wang2023largelanguagemodelsfair, chen2024humansllmsjudgestudy}.

\subsection{LLMs for code generation}

While LLMs are mainly focused on natural language generation, a significant effort has also been made to \textbf{generate programming code} with them. Being expressed as written text, programming languages seem like a natural extension of human languages. However, although the syntax underlying programming languages is way simpler and more restricted than that of natural languages, they do not allow for deviations, and thus are less fault-tolerant than human languages. Furthermore, a program is generated with the purpose of performing a particular function, so even syntactically correct programs can be of no use if they do not fulfill their target function.

Similarly to how natural languages are dealt with, a number of high-performing LLMs have been trained with texts both in natural languages and programming languages \cite{gpt-4, llama, bloom, stablelm2, qwen2}, which grants them noticeable abilities in code generation. Other LLMs are highly specialized in code generation, either by fine-tuning a base model with a code corpus (\textbf{Codex} \cite{chen2021evaluatinglargelanguagemodels}), or by training from scratch using codebases and related technical sources (\textbf{StarCoder} \cite{starcoder}, \textbf{StarCoder2} \cite{starcoder2andthestack2}, \textbf{StableCode} \cite{stablecode}, \textbf{DeepSeekCoder} \cite{deepseekcoder}).

Several useful tasks can be addressed with these kind of LLMs: generating the body of a function given a description in natural language, writing documentation or explanations for a given piece of code, producing unit tests for a function, reviewing pull requests or proposing patches to solve issues, among others. In this paper we will focus on the task of \textbf{generating a function given a natural language description and a set of sample unit tests}.

\subsection{Evaluating LLMs in code generation tasks}

Software development is a broad field that encompasses multiple aspects. While requiring some shared basic knowledge and skills, programming a web application, an AI algorithm, or a video game are very different tasks. This is not only due to the differences in programming languages but also because of the variety of tools and standards employed in each domain. Moreover, developing a complete software solution also requires the ability to design and understand systems composed of various modules, to collaborate effectively with other project members, and to interpret user requirements and translate them into code.

A number of previous works have analyzed the problem of evaluating LLMs on coding tasks, or more generally, evaluating the performance of synthetic code generators, which has been an active line of research before the advent of LLMs \cite{gulwaniPS17}. This section aims to provide a review of relevant prior research on this topic.

\textbf{code-docstring-corpus} \cite{barone2017parallelcorpuspythonfunctions} is a dataset of 161,630 pairs of function docstrings and corresponding code in Python,  extracted by scrapping projects from GitHub \cite{github}. Baseline models for generating code from the docstring and viceversa are built using Neural Machine Translation models, whose performance is then evaluated using the metric \textbf{BLEU} \cite{bleu}. BLEU is a metric widely used for automated machine translation, which boils down to measuring the number of matching n-grams between the generated text and a reference solution, penalizing also for brevity.

The \textbf{CodeXGLUE} dataset \cite{lu2021codexgluemachinelearningbenchmark} attempts to mimic the famous GLUE dataset \cite{glue} for Natural Language Processing, aiming to create a reference benchmark for a set of diverse tasks related with automated code generation. To this end, a number of previously existing datasets are aggregated, for tasks such as clone detection, defect detection, code repair, code translation, code search, and text-to-code generation among others. The task of text-to-code generation is based on the CONCODE dataset \cite{iyer2018mappinglanguagecodeprogrammatic}, an extraction from around 33,000 Java projects on GitHub that results in around 104,000 examples consisting of natural language descriptions, code environments and code snippets, with the objective of generating class member functions from Javadoc-style comments. The evaluation metrics proposed for the task are exact match, BLEU and \textbf{CodeBLEU} \cite{codebleu} scores. CodeBLEU is an adaption of the BLEU metric for programming languages. It takes into account the level of match of the generated and reference solutions not only in terms of n-grams, but also using the Abstract Syntax Trees (AST) underlying the code, and the data-flows representing the semantics on how each variable is computed from other variables. 

In \textbf{Search-based Pseudocode to Code (SPoC)} \cite{NEURIPS2019_7298332f} the authors address the task of translating pseudocode to actual programming code, scraping 18,356 solutions of 677 programming problems from Codeforces \cite{codeforces}, then hiring a team 59 crowdworkers to translate them to pseudocode. A sequence to sequence LSTM model is trained over a subset of the data, and solutions are generated for a separate test set. A solution is deemed as valid if it successfully compiles and passes a set unit tests. To measure the performance they propose the \textbf{success rate at B} metric, defined as the fraction of test problems where the system generates an accepted program under the budget of $B$ trials. The authors argue this metric is superior to BLEU, as the latter fails to account for functional correctness and only focuses on the surface form of the code.

\textbf{Automated Programming Progress Standard (\textbf{APPS})} \cite{hendrycks2021measuringcodingchallengecompetence} addresses the more ambitious task of generating code from descriptions in natural language. It consists of 10,000 coding problems together with unit tests and ground-truth solutions written by humans. The data is obtained by scrapping several websites centered around coding challenges (AtCoder, CodeChef, Codeforces, Codewars,
HackerRank, Kattis, and LeetCode), followed by a manual curation process. Coding challenges are also arranged according to their difficulty into three groups: introductory level, coding interview level, and competition level. GPT-2 \cite{gpt-2} and GPT-Neo \cite{gpt-neo} models are fine-tuned and tested over this dataset,  as well as a few-shot prompting approach with GPT-3 \cite{gpt-3}. Two evaluation metrics are proposed: an average number of unit tests completed, and a \textit{strict} metric that requires a solution to pass all unit tests. The evaluation results show that interview-level exercises are indeed harder to solve by the LLMs, and not a single competition-level problem can be solved successfully when using the \textit{strict} metric.

\textbf{HumanEval} \cite{chen2021evaluatinglargelanguagemodels} is another evaluation dataset presented along with Codex, a GPT-3 \cite{gpt-3} model fine-tuned over GitHub data. The dataset consists of hand-written Python function signatures and docstrings from which the LLM is expected to produce a working implementation. HumanEval contains 164 programming problems, covering tasks such as language comprehension, reasoning, algorithms and simple mathematics, of a difficulty at the level of easy code interviews. The authors emphasize the relevance of the problems being generated from scratch for their work, as they find that public repositories already contain solutions for problems posted at Codeforces, which are part of other evaluation datasets such as the aforementioned APPS dataset \cite{hendrycks2021measuringcodingchallengecompetence}. Regarding the metrics, the authors also introduce the \textbf{pass@k} metric, which is similar to the \textit{success rate at $B$} metric introduced before, the main difference being that a number $n > k$ of samples is generated, and the extra samples are used to compute an estimate with reduced variance. When analyzing the performance of Codex, it is found that albeit being effective, a "strong student who completes an introductory computer science course is expected to be able to solve a larger fraction of problems than Codex-12B". The authors also find that the quality of the generated code decreases as the number of logical steps to be performed increases.

A number of \textbf{extensions of HumanEval} have also been proposed. \textbf{HumanEval-X} \cite{zheng2023codegeex} manually translates the programming problems from HumanEval into 4 different programming languages (C++, Java, JavaScript, and Go), resulting in a total of 820 human-crafted data samples together with test cases. \textbf{HumanEval-XL} \cite{peng-etal-2024-humaneval-xl} extends this further not only in terms of programming languages (12) but also in the number of natural languages used in the docstrings (23), comprising of a collection of 22,080 prompts. The prompts in natural languages other than English are obtained through machine translation with GPT-4 \cite{gpt-4}. \textbf{EvalPlus} \cite{liu2023codegeneratedchatgptreally} augments the quantity of unit tests of HumanEval by a factor of 80, by making use of ChatGPT and type-aware mutations of the test inputs. This greater coverage in tests leads to a noticeable decrease in performance of many LLMs, revealing the LLM-generated code is not as good as presented by other benchmarks. Also, vague descriptions of some problems were improved, and incorrect ground-truth solutions (about $10\%$ of the dataset) were manually corrected. On a more recent version of this benchmark, the authors also include an extended version of MBPP, increasing its number of tests by a factor of 35. A leaderboard of publicly available at \url{https://evalplus.github.io/leaderboard.html}.

\textbf{Mostly Basic Python Programs (MBPP)} \cite{austin2021programsynthesislargelanguage} is, as its name implies, a dataset of 974 simple python functions, together with unit tests. As input to the model a description in natural language is provided, together with 3 input-output examples in the form of asserts showing the expected behavior of the function to generate. The problems are all designed to be solvable by entry-level human programmers. They were manually created through crowdsourcing, with a subset of 426 problems being manually reviewed and edited by the authors. Performance is measured using the  \textit{pass@$k$} metric, as the authors find BLEU does not correlate well with synthesis performance. It is also found that the tested LLMs struggle to generate valid solutions when the problem requires solving several subproblems, and when the problem is a modification of a popular problem which is present in the training data with high probability. Better performance is observed over the subset of curated problems. Overall, the problems solved by the LLM are the shortest and simplest in the dataset, and about 80 generation samples are required to obtain 1 or 2 valid solutions. Improvements in performance can be attained by allowing the LLM to establish a multi-turn dialogue with a human expert that suggests corrections in natural language. The authors also test whether the LLM is able to predict the output of a given program, which results in a poor performance. Interestingly, the performance is increased if a natural language description of the program is presented, which strongly suggests the LLM is unable to actually "run" the code to predict its output, something a human programmer is able to do manually for small programs. The authors conclude that we are a long way from generating models that can synthesize complex applications without human supervision.

\textbf{Open-Domain EXecution-Based dataset  (ODEX)} \cite{wang-etal-2023-execution} strives to present the LLM with programming tasks better aligned with those faced by developers, moving beyond competition-like challenges or problems that can be solved with native Python funcions. 945 pairs of natural language requests and corresponding codes were gathered from Stack Overflow crawls \cite{stackoverflow}, spanning 4 natural languages and the usage of 79 Python libraries. Unit tests were created manually by a team of undergraduate students. \textit{pass@k} is used as the main evaluation metric, although other metrics that do not require unit-testing are also checked, including BLEU and CodeBLEU; however, the authors conclude these metrics do not correlate well with \textit{pass@k}. Evaluation over two state-of-the-art LLMs for coding shows that performance seems to be better for those coding tasks involving libraries frequently used on GitHub ($\sim$50\% pass@1), although frequent but complex to use libraries such as matplotlib or tensorflow show poor performance ($<$10\% pass@1). The authors also observe that injecting one input-output example in the prompt significantly improves accuracy.

\textbf{BigCodeBench} \cite{zhuo2024bigcodebenchbenchmarkingcodegeneration} also aims to test the LLM with programming tasks similar to actual work performed by developers in real projects, by requiring the solutions to use function calls to libraries from 7 different domains (computation, visualization, networking, ...). Making use of seed examples from the ODEX benchmark, a group of human annotators working in an interactive loop with GPT-4 \cite{gpt-4} generated 1140 problems that require using a total of 139 libraries. Two versions of the benchmark are produced: BigCodeBench-Complete, which prompts the LLM with well-structured docstrings, and BigCodeBench-Instruct, in which natural language descriptions are provided. Through an evaluation of several LLMs the authors find that even the best models can only obtain a $60\%$ solve rate at BigCodeBench-Complete, and a $50\%$ at BigCodeBench-Instruct, while the human performance is $97\%$; this shows LLMs have still a long way to go in terms of understanding and using several tools jointly. The authors also observe that the performance varies between domains, something they attribute to the training data used. A leaderboard with the benchmark results is available at \url{https://bigcode-bench.github.io/}.

Continuing this trend, \textbf{SWE-bench} \cite{jimenez2024swebenchlanguagemodelsresolve, swebenchwebsite} is perhaps the benchmark that most closely reflects the actual tasks carried out by a professional developer. The benchmarks consists of 2294 issues and their corresponding pull requests solving them, scraped from popular repositories on GitHub. The LLM is asked to generate a code patch that solves the issue, given its description and the entire codebase of the repository. As evaluation metric, the proposed patch is applied to the codebase and the repository tests are run against the resultant code, checking whether all tests are successful. All the pull requests selected for the benchmark include changes in the tests that check whether the proposed solution actually solves the issue, therefore the LLM cannot just produce a null patch to pass all the tests. This benchmark is remarkably hard for LLMs due to their inherent context length limit: codebases have 438K lines of code on average, and a correct patch generally involves modifying several code files. Thus, any performant solution must integrate the LLM as a component of a larger framework that includes retrieval strategies. The authors of the benchmark resort to a basic retrieval method (BM25 \cite{bm25}) to select a reduced subset of the code files, which are then fed as context to the LLM, solving only 1.96\% of the issues with a Claude-2 model. More advanced approaches have attained up to 22.06\% solve rate at the date of writing.


\textbf{SWE-bench verified} \cite{swebenchverified} is a human-validated subset of SWE-bench, where a team of professional developers have carried out an annotation job, classifying each task according to how well-defined it is in terms of description and unit tests, and how long would take a human developer to solve it.  Through this procedure those tasks that are poorly defined are removed, producing a fairer evaluation benchmark. Benchmark results show that competing models perform significantly better on this reduced subset, proving the usefulness of well-defined tasks for fair evaluations.

\textbf{LMSYS Chatbot Arena Leaderboard} \cite{chiang2024chatbotarenaopenplatform} presents an alternative method of benchmarking, where the performance of more than a hundred LLMs is compared through the evaluations of human volunteers. The Leaderboard offers an open access to a web interface where any visitor can make a request in natural language, and the response from two different LLMs are returned. The user can then vote a preferred answer, and the system maintains en Elo rating based on these votings. Although the Leaderboard aims to measure the performance of LLMs under general-purpose requests, it also includes a specialized ranking for coding tasks.











\section{Measuring software development capabilities through programming exercises}
\label{sec:measuringSWcapabilites}


In this paper we will limit ourselves here to measuring one specific aspect of software development: \textbf{solving programming exercises}, also popularly known as code challenges or "katas". In these exercises, the programmer is presented with a description of a specific functionality to implement, such as creating a function that determines whether a given number is even or odd. The programmer must then develop the code to implement this functionality, following instructions regarding the programming language to use and the format to follow.

The advantage of this type of exercise is that it is possible to automatically verify whether the code produced is correct. The kata designer creates a series of unit tests that challenge the programmer's solution, verifying whether the outputs produced by the code meet expectations across various scenarios. The performance of an LLM can then be measured with already established metrics such as \textit{pass@k}.

\begin{figure}
    \centering
    \includegraphics[width=1\textwidth]{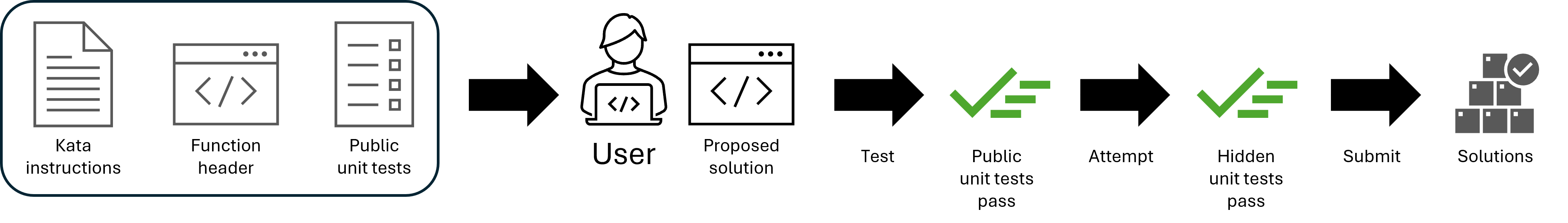}
    \caption{Flow of kata completion by a user in Codewars. The user receives the information of the kata in the form a description, a function header or template to follow in the solution, and a set of public unit tests. When user proposes a solution, it is first checked against the public unit tests, and then again a larger set of hidden tests. If the proposed solution succeeds both checks, it is accepted and added to a repository of valid solutions.}
    \label{fig:codewarsFlow}
\end{figure}

An instance of a software development community that makes use of this entire system of katas is Codewars \cite{codewars}.  Figure \ref{fig:codewarsFlow}  presents the flow followed by a user in order to successfully complete a kata. A relevant fact is that the proposed solution is tested against two sets of units tests, one of them kept hidden from the user. In this way, the user cannot create a trivial solution that just returns the expected outputs for the unit tests. When the user completes all the tests, the proposed solution is added to a repository of all valid solutions created for the kata so far; this repository is kept hidden from all users except those who have managed to submit a valid solution, or those who have decided to forfeit the exercise, losing their chance to complete it successfully.

In addition to containing thousands of katas created by its members in a wide range of programming languages, Codewars also includes a system for ranking katas by difficulty, modeled after the kyus, which stand for experience levels in martial arts. Katas are categorized from kyu 8 (easiest) to kyu 1 (most difficult):

\begin{itemize}
    \item \textbf{Kyu 8 and 7}: Basic exercises that require knowledge of the fundamentals of a programming language and the ability to solve simple tasks with them. Examples include determining whether a number is even or sorting the digits of a number. These are appropriate for someone learning to program for the first time, but trivial for an experienced developer.
    \item \textbf{Kyu 6 and 5}: Intermediate exercises that require more detailed knowledge of the programming language and familiarity with basic data structures and algorithms.
    \item \textbf{Kyu 4 and 3}: Advanced exercises, most of which require knowledge of more complex algorithms, mathematical foundations, or the implementation of multiple subroutines.
    \item \textbf{Kyu 2 and 1}: Highly complex exercises. These may involve implementing the internal logic of what could be a complete application, such as a perfect Minesweeper player or a symbolic differentiation system. In other cases they require non-obvious mathematical developments or the application of code optimization techniques to meet strict runtime or code length requirements, such as calculating the area of intersection between two circles with a function shorter than 128 characters.
\end{itemize}

\begin{figure}
    \centering
    \includegraphics[width=1\textwidth]{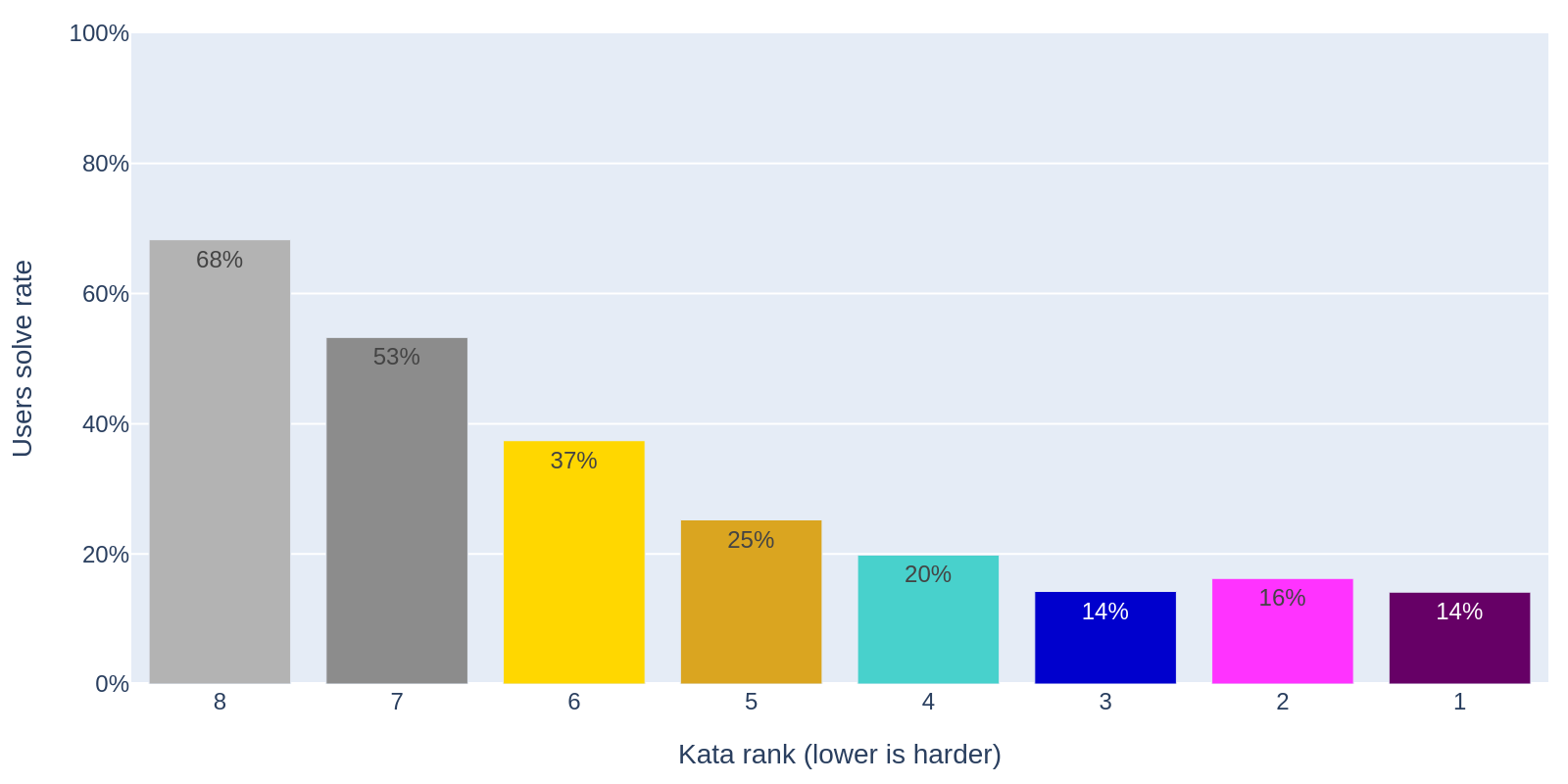}
    \caption{Percentage of users who manage to complete a kata, once they decide to start it, according to its difficulty level, regardless of the programming language used.}
    \label{fig:usersRank}
\end{figure}

Analyzing the statistics of approximately 8,000 Codewars katas (Figure \ref{fig:usersRank}), we can confirm that, in general terms, the kyu level correlates with the actual difficulty: the more complex katas are successfully completed by a smaller proportion of the users who attempt them.

\section{Experimental setup}

For this study, we have chosen to use GPT-4o-mini (version 2024-07-18)  \cite{gpt-4o} as the LLM to test, given that, at the time of writing, this language model ranks \#3 for general tasks and \#2 for programming tasks in the LMSYS Chatbot Arena Leaderboard \cite{chiang2024chatbotarenaopenplatform}. Although the full model (GPT-4o) demonstrates better performance in this ranking, the fact that the mini version is approximately 30 times cheaper makes it a more practical solution.

\begin{figure}
    \centering
    \includegraphics[width=0.9\textwidth]{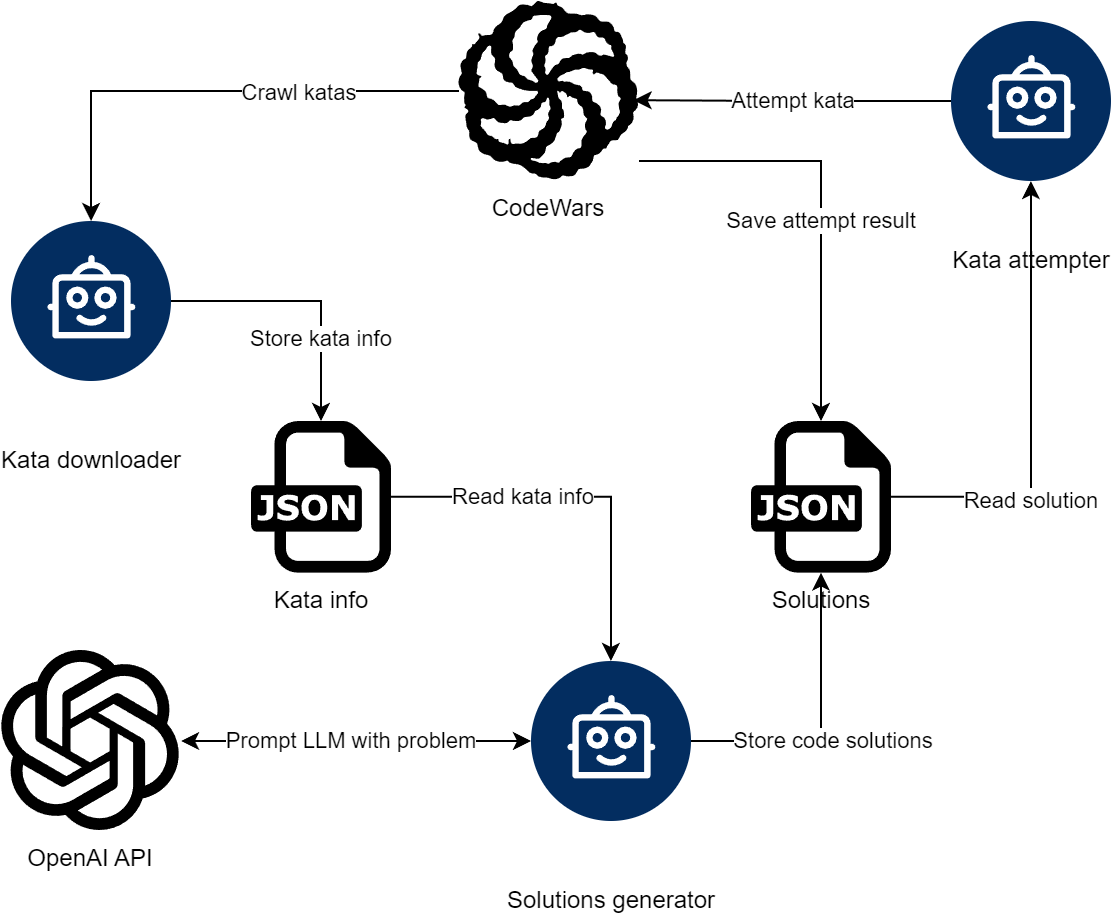}
    \caption{Network of bots used to download kata information, generate solutions using an OpenAI model, and verify whether the solutions are correct.}
    \label{fig:botDiagram}
\end{figure}

Of course, a language model cannot perform this task on its own accord. Therefore, for this study a series of bots that interact between Codewars and the OpenAI API were developed, whose structure is show in Figure \ref{fig:botDiagram}. The entire system implementation was done in Python, using Selenium \cite{selenium} to create the bots that interact with Codewars.

It should be noted that the Codewars code of conduct prohibits the use of AI systems to solve problems in an automated manner. Therefore, the bots deployed in this study used Codewars accounts that were created solely for this task and were subsequently deleted, ensuring that the rankings of real users, who rely on this platform to measure their programming skills, were not affected. Similarly, all crawling was performed responsibly to avoid causing service disruptions.

\begin{figure}
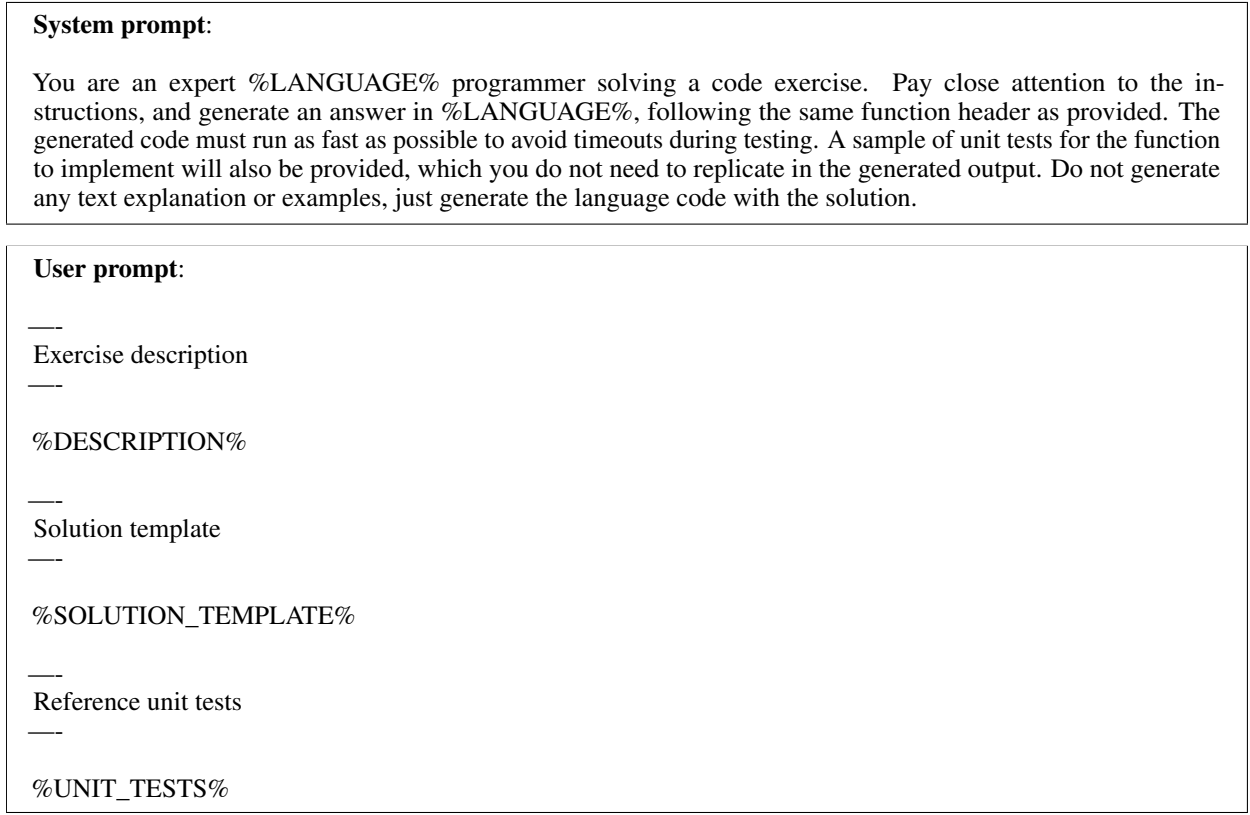

\begin{mdframed}
\textbf{System prompt}: \\
\\
You are an expert \%LANGUAGE\% programmer solving a code exercise. Pay close attention to the instructions, and generate an answer in \%LANGUAGE\%, following the same function header as provided. The generated code must run as fast as possible to avoid timeouts during testing. A sample of unit tests for the function to implement will also be provided, which you do not need to replicate in the generated output. Do not generate any text explanation or examples, just generate the {language} code with the solution.
\end{mdframed}
\begin{mdframed}
\textbf{User prompt}: \\
\\
----\\
Exercise description \\
----\\
\\
\%DESCRIPTION\% \\
\\
----\\
Solution template \\
----\\
\\
\%SOLUTION\_TEMPLATE\% \\
\\
----\\
Reference unit tests \\
----\\
\\
\%UNIT\_TESTS\%
\end{mdframed}
\caption{Prompts used for the coding solutions generation.}
\label{fig:prompts}
\end{figure}

When interacting with the LLM the prompts shown in Figure \ref{fig:prompts} were used, where the following relevant information is injected: target programming language in which to develop the solution, description of the task, template that the solution must follow, and public unit tests visible to the user. In this way the LLM has the same information to solve the problem as a human user would.

Each kata in Codewars may accept solutions written in a range of programming languages. While the majority of the katas are only designed for a few languages, the most popular katas may raise this number up to 50. For this study we have decided to focus on a subset of 8 programming languages, being:

\begin{itemize}
    \item \textbf{JavaScript} \cite{javascript}: used mainly for web development, JavaScript has been the most widely used programming language on GitHub for the last 4 years \cite{GitHub_GitHub_Innovation_Graph_2024}.
    \item \textbf{Python} \cite{python}: used for a wide range of applications, including back-end development, data science and machine learning. It is the \#2 most used language at the time of writing, according to GitHub statistics.
    \item \textbf{C} \cite{kernighan2006c}: a general purpose language created in 1970, C continues to be a major cornerstone in many industrial applications, and is the backbone of many compute-efficient libraries in high-level languages such as Python. Currently holds the \#8 position in the ranking of most used languages.
    \item \textbf{R} \cite{rlang}: although its usage has decreased significantly over the last years (currently \#26), it is still an alternative to Python for statistics and data science purposes.
    \item \textbf{Julia} \cite{julia}: yet another Python competitor, still not frequently used (not registered in GitHub top 50), but active in some areas of science.
    \item \textbf{Solidity} \cite{solidity}: a specialized programming language for implementing smart contracts over the Ethereum blockchain. Solidity has gained popularity in the recent years, being currently the \#36 programming language on GitHub.
    \item \textbf{Fortran} \cite{fortran}: first proposed in 1957, Fortran is still actively used nowadays to implement highly-efficient numerical computations in the domain of High Performance Computing. The \textit{de facto} standard libraries for linear algebra BLAS \cite{blas} and LAPACK \cite{lapack99} are implemented in Fortran. It has ranked in positions \#49 or \#50 for 5 quarters across the last 4 years.
    \item \textbf{COBOL} \cite{cobol}: a proprietary language created by IBM in 1959, in the current age is mainly used in legacy systems in the banking and public sectors. It is estimated that 43\% of the banking systems are built on COBOL, and 95\% of the code involved in ATM swipes transactions is also written in COBOL \cite{cobolusage}.
\end{itemize}

With this subset of languages we aim to cover a selection of popular, niche and legacy languages. Popular languages are those for which a larger number of software developers would benefit from a coding LLM. Conversely, in the case of less common languages fewer human experts exist, and thus an "LLM-expert" would be of great help. This is especially true for legacy languages, which currently pose a significant maintainability threat: the average age of COBOL programmers is 45-55 years old \cite{cobolusage}, which will produce a lack of active experts in the years to follow.

\begin{figure}
    \centering
    \includegraphics[width=1\textwidth]{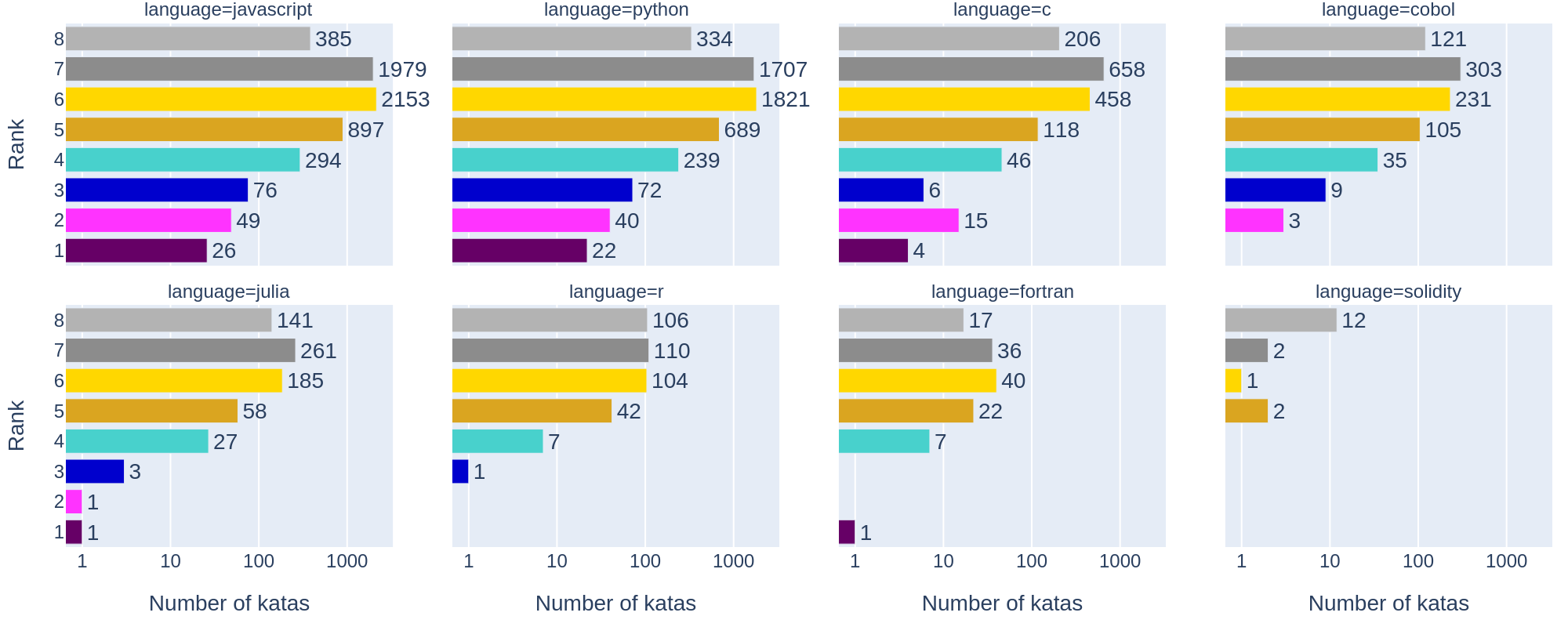}
    \caption{Number of katas processed in this work, divided by programming language and difficulty (rank). Lower rank means higher difficulty. Some combinations of languages and rank have no katas available in Codewars.}
    \label{fig:processedKatas}
\end{figure}

Figure \ref{fig:processedKatas} presents the distribution of katas processed in this work, organized by language and difficulty rank. These amount to the whole set of katas available in Codewars for the selected languages at the time of writing (14,346), except for some katas that were discarded due to incompatibilities in the code generated by the LLM and the injection mechanism implemented in the "Kata attempter bot": the solution to these katas requires generating some complex regular expressions, which contain special symbols that raise an exception in the Selenium JavaScript interpreter when injecting the generated code in the Codewars webpage. The discarded katas amount to 57, just a 0.4\% of the total number available; thus this loss of katas is negligible for this work.

It must also be noted that the distribution of katas across languages and difficulty ranks is not uniform. As one might expect, the number of available katas is loosely correlated to the popularity of the language. Regarding the rank, most katas belong to low or intermediate kyus. The hardest katas are scarce and mainly available for the most popular languages.

\subsection{Codewars Score}

\begin{figure}
    \centering
    \includegraphics[width=1\textwidth]{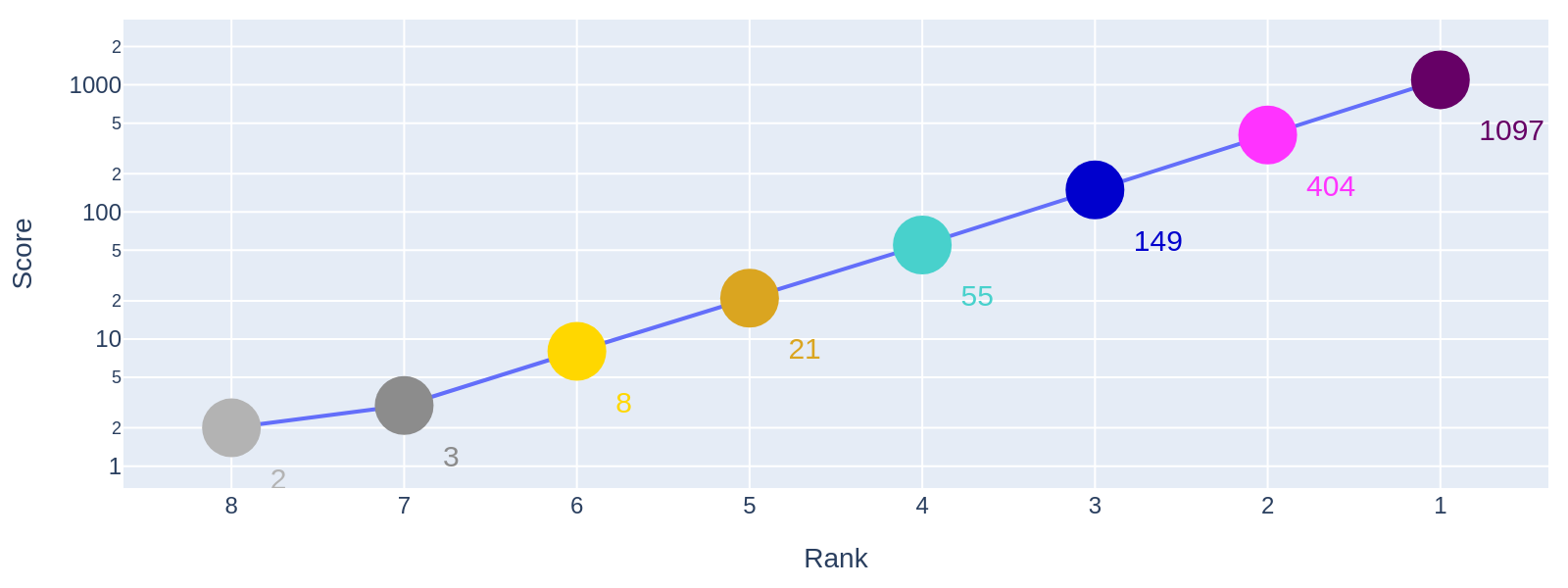}
    \caption{Score awarded in Codewars for successfully completing a kata of a given rank. Lower ranks represent harder katas.}
    \label{fig:codeWarsScoresLine}
\end{figure}

The Codewars community makes use of a custom metric to compute the overall skill level of software developers, taking into account the effort involved in solving katas of different ranks. Regardless of the programming language, a successful solution to a kata is awarded a number of points that depend on the rank or kyu of the solved kata. The number of points awarded closely follows an exponential function of the kata rank, as shown in Figure \ref{fig:codeWarsScoresLine}. This is a fair scoring method, as harder katas require a much deeper degree of knowledge and involvement on the side of the developer. Under this scoring scheme, a user has a total score that is the sum of the scores attained from each solved kata.

In this paper we shall adopt a similar scoring method to measure the performance of the LLM. However, and as opposed to humans, the bot network developed can propose a solution to a kata in a matter of seconds, regardless of its difficulty. While this superhuman coding speed is a merit on its own accord, the focus of this paper is to measure the capabilities of the LLM in terms of being able to find solutions to katas of varying complexities. Therefore, we shall use a modified version of the Codewars score that measures an average performance of the LLM across all katas, using the expression

\begin{equation}
    \label{eq:codeWarsScore}
    p_l = \sum_{i=1}^8 s_i r_{li},
\end{equation}

where $p_l$ is the computed performance of the LLM for programming language $l$, $s_i$ is the Codewars score granted for a kata of rank $i$ (following Figure \ref{fig:codeWarsScoresLine}), and $r_{li}$ is the ratio of katas of rank $i$ that the LLM is able to solve in language $l$. In summary, the metric we shall use is a weighted average of the frequency of the LLM success in solving exercise, putting more weight in harder tasks. A higher value on this metrics means the LLM is able to solve tasks of higher complexity more frequently.

\section{Results}

\subsection{Overall performance}

\begin{figure}
    \centering
    \includegraphics[width=1\textwidth]{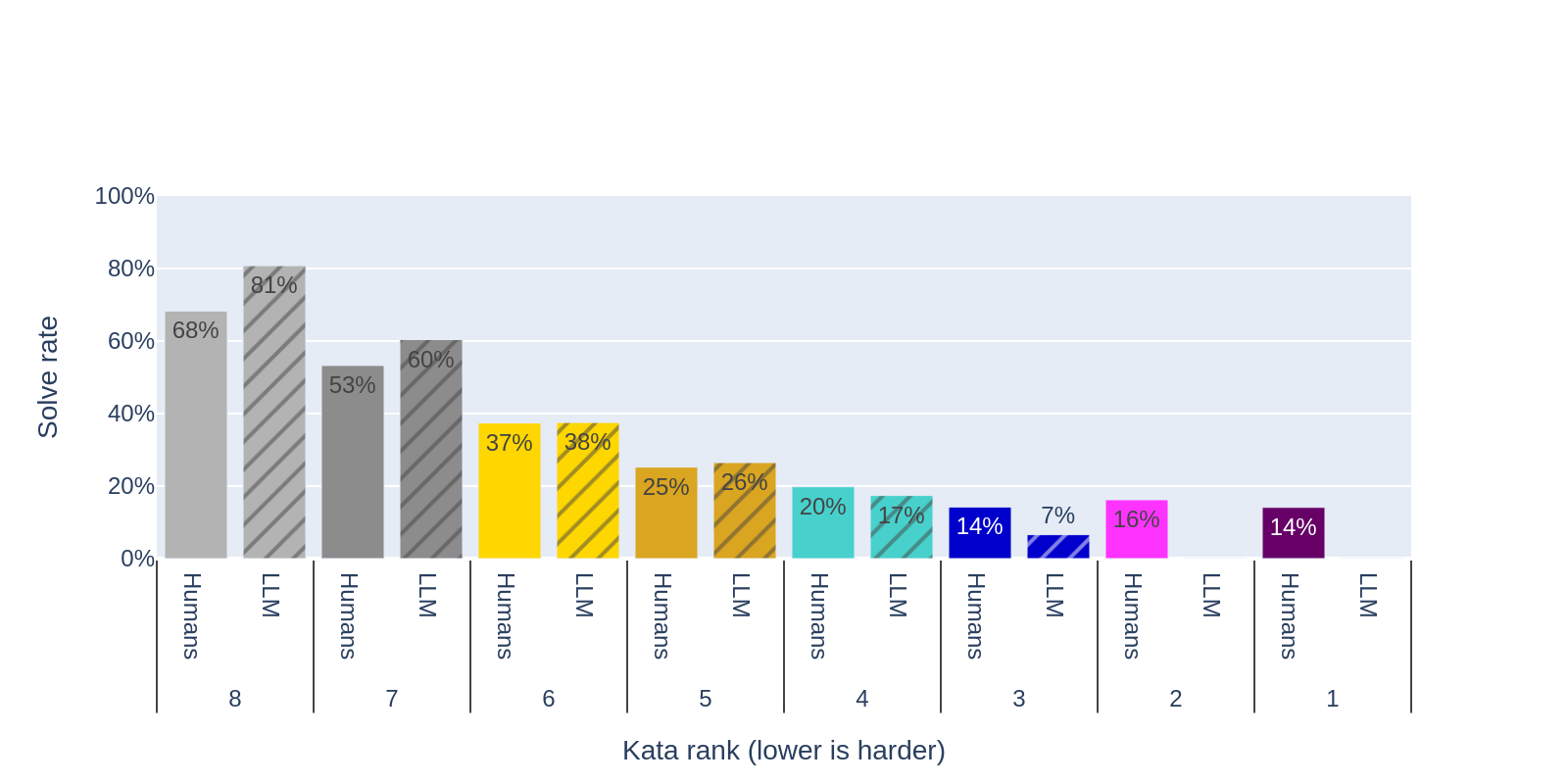}
    \caption{Average percentage of users who successfully complete a kata, and the percentage of katas successfully completed by the LLM, categorized by difficulty level, regardless of the programming language used.}
    \label{fig:usersLLmRank}
\end{figure}

Figure \ref{fig:usersLLmRank} shows the performance levels achieved by the LLM in comparison to human developers. Several points stand out:

\begin{itemize}
    \item \textbf{The LLM outperforms humans in solving easy tasks} (ranks 8 and 7).
    \item \textbf{The performance of humans and the LLM is similar when solving intermediate tasks} (ranks 6 and 5) and even some early advanced tasks (rank 4).
    \item However, when reaching rank 3 tasks, the LLM's performance significantly declines. Furthermore, \textbf{for the most complex exercises} (ranks 2 and 1), \textbf{the LLM fails completely}, unable to solve any of them. This is in accordance with previous benchmark evaluations \cite{hendrycks2021measuringcodingchallengecompetence, austin2021programsynthesislargelanguage, swebenchverified}, where the evaluated LLMs failed to solve any of the problems with the highest difficulty level.
\end{itemize}

\subsection{Performance by programming language}

\begin{figure}
    \centering
    \includegraphics[width=1\textwidth]{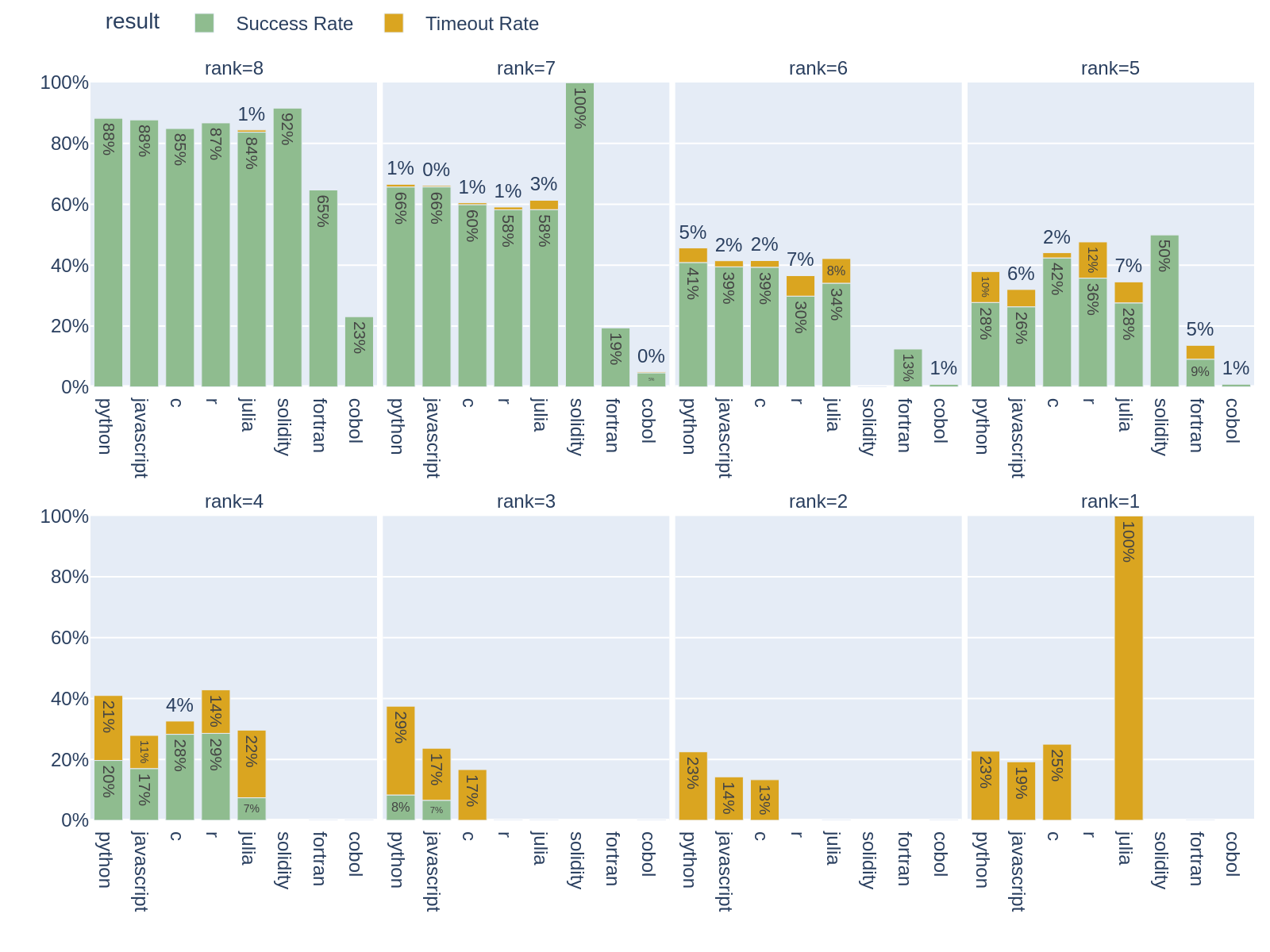}
    \caption{Percentage of katas successfully complete by the LLM, according to programming language and difficulty rank. The green bars mark the percentages of tasks successfully completed, while yellow bar represent the percentage of tasks that failed due to the generated code exceeding the time limit (Timeout).}
    \label{fig:rankLanguages}
\end{figure}

Figure \ref{fig:rankLanguages} deepens the analysis by examining the LLM's performance in solving katas of various difficulty levels across different programming languages.  It is noteworthy that the LLM's performance significantly depends on the programming language used:

\begin{itemize}
    \item \textbf{The LLM performs better in widely-used languages}: \textbf{JavaScript} and \textbf{Python} are the two most commonly used general-purpose languages on GitHub. Additionally, in these languages, the proposed solutions for more difficult exercises do not seem to be entirely incorrect; rather, they often fail due to insufficient optimization (Timeout).
    \item Even in less commonly used languages, such as \textbf{C} or \textbf{R} , \textbf{performance remains fairly strong}, although the LLM does not manage to solve any rank 3 katas in these languages (6 available for C, 1 for R).
    \item In languages like \textbf{Julia}, which are even less widely adopted, the LLM seems to fare reasonably well. Although performance is lower than in C or R for intermediate or advanced exercises, it is reliable for basic tasks. A similar trend is observed with Solidity.
    \item In legacy languages (\textbf{Fortran} and \textbf{COBOL}), \textbf{performance drops dramatically}. COBOL, in particular, shows catastrophic results, and visual inspection of the generated code reveals many cases of syntax errors, something extremely rare for other languages.
\end{itemize}

\begin{figure}
    \centering
    \includegraphics[width=1\textwidth]{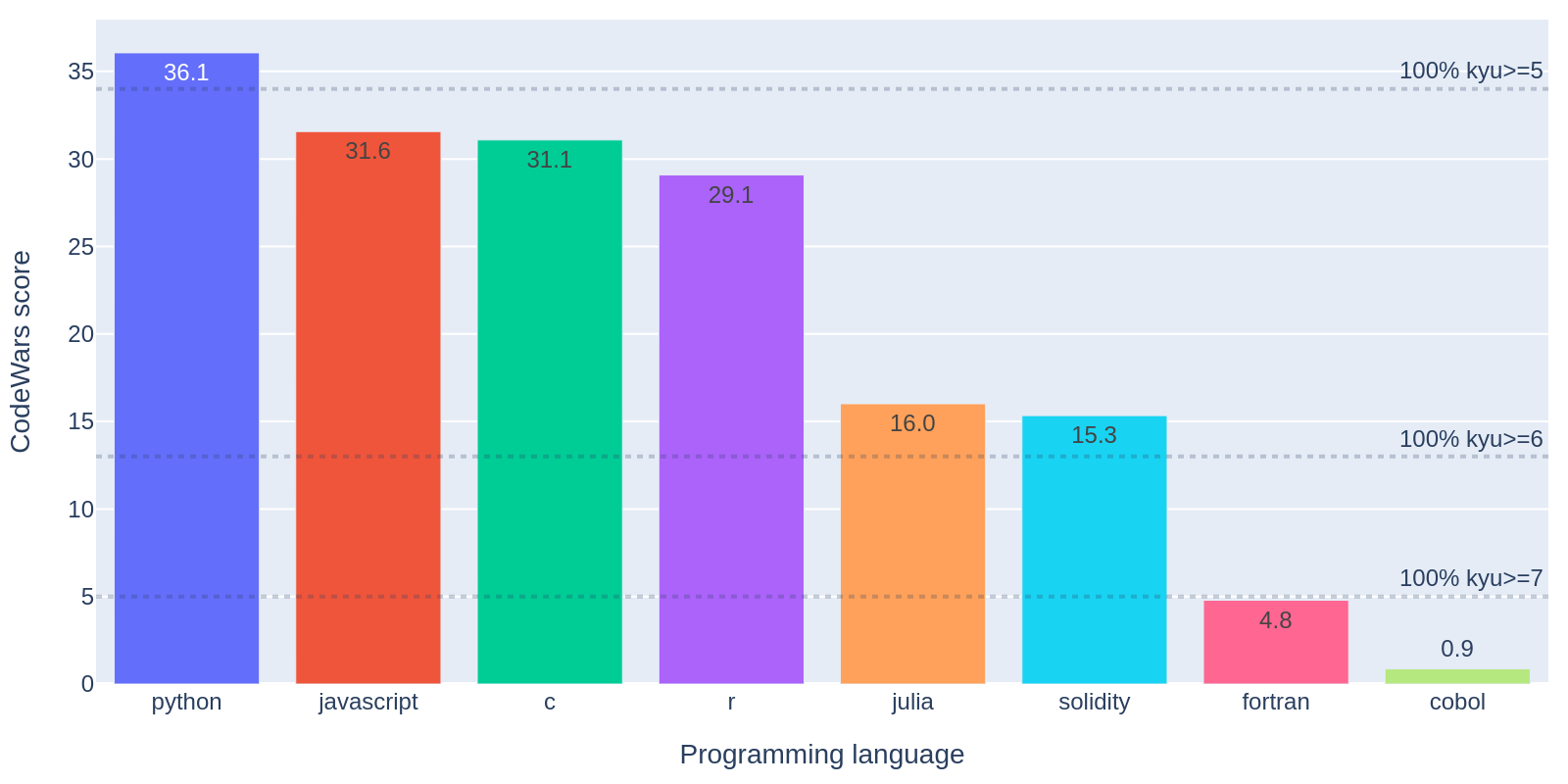}
    \caption{Performance levels of the LLM programming in different languages, measured using the Codewars score. The score levels that would be obtained if 100\% of the katas of a given kyu or higher were successfully completed are also indicated.}
    \label{fig:languagesCodewarsSCores}
\end{figure}

To compute an overall performance by language, we make use of the Codewars score defined before in Equation \ref{eq:codeWarsScore}. This leads to the scores represented in Figure \ref{fig:languagesCodewarsSCores}. \textbf{Python} emerges as the language in which the LLM performs best. However, a score of 36.1 indicates that it is still far from consistently providing valid solutions for advanced exercises: a system that solved all katas up to kyu 4 or higher would have achieved 87 points.

\subsection{Factors motivating performance differences across languages}

\begin{figure}
    \centering
    \includegraphics[width=1\textwidth]{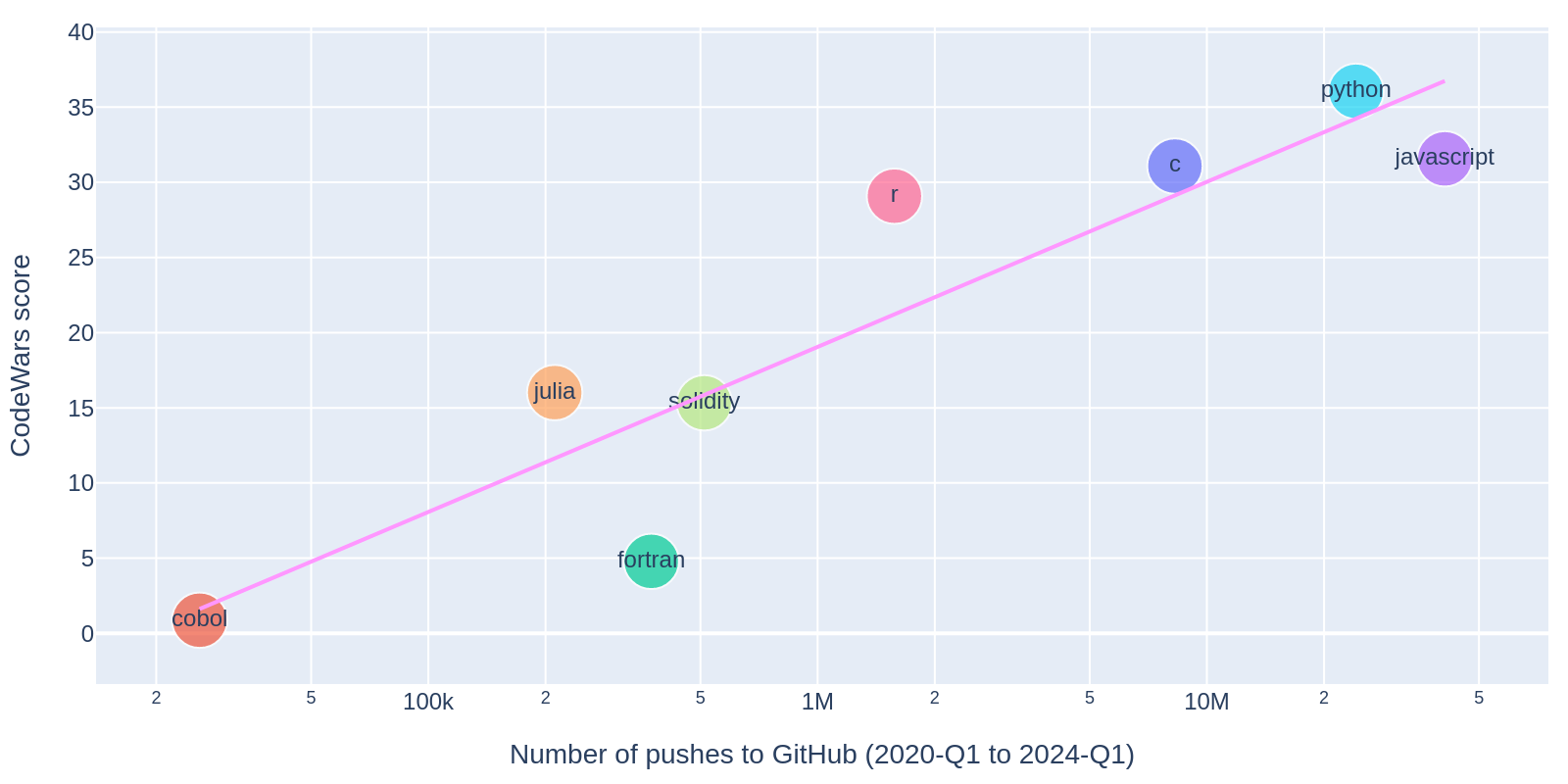}
    \caption{Performance levels of the LLM programming in different languages, measured by the Codewars score, compared to the number of code pushes in different languages on GitHub.}
    \label{fig:codeWarsScoresGithubPushes}
\end{figure}

Previous research has already observed that other LLMs trained on public GitHub repositories perform especially well when coding in popular programming languages such as Python or Java \cite{zheng2023codegeex}. The training data used for GPT models is not publicly known, but it is reasonable to assume that GitHub repositories have been used as well. To the best of our knowledge, there are no public statistics on the total lines of code per language stored in this community, but there are data on the number of pushes to repositories, categorized by language \cite{GitHub_GitHub_Innovation_Graph_2024}. From these, we can generate the analysis shown in Figure \ref{fig:codeWarsScoresGithubPushes}.

A clear correlation appears: not surprisingly, \textbf{programming languages with more code available for training are the ones in which the LLM can perform better}. This is good news for programmers working in the most widely-used languages, but disappointing for those considering using this technology to maintain or adapt the vast COBOL and Fortran codebases that still exist in legacy production systems.

\subsection{Influence of kata age}

\begin{figure}
    \centering
    \includegraphics[width=1\textwidth]{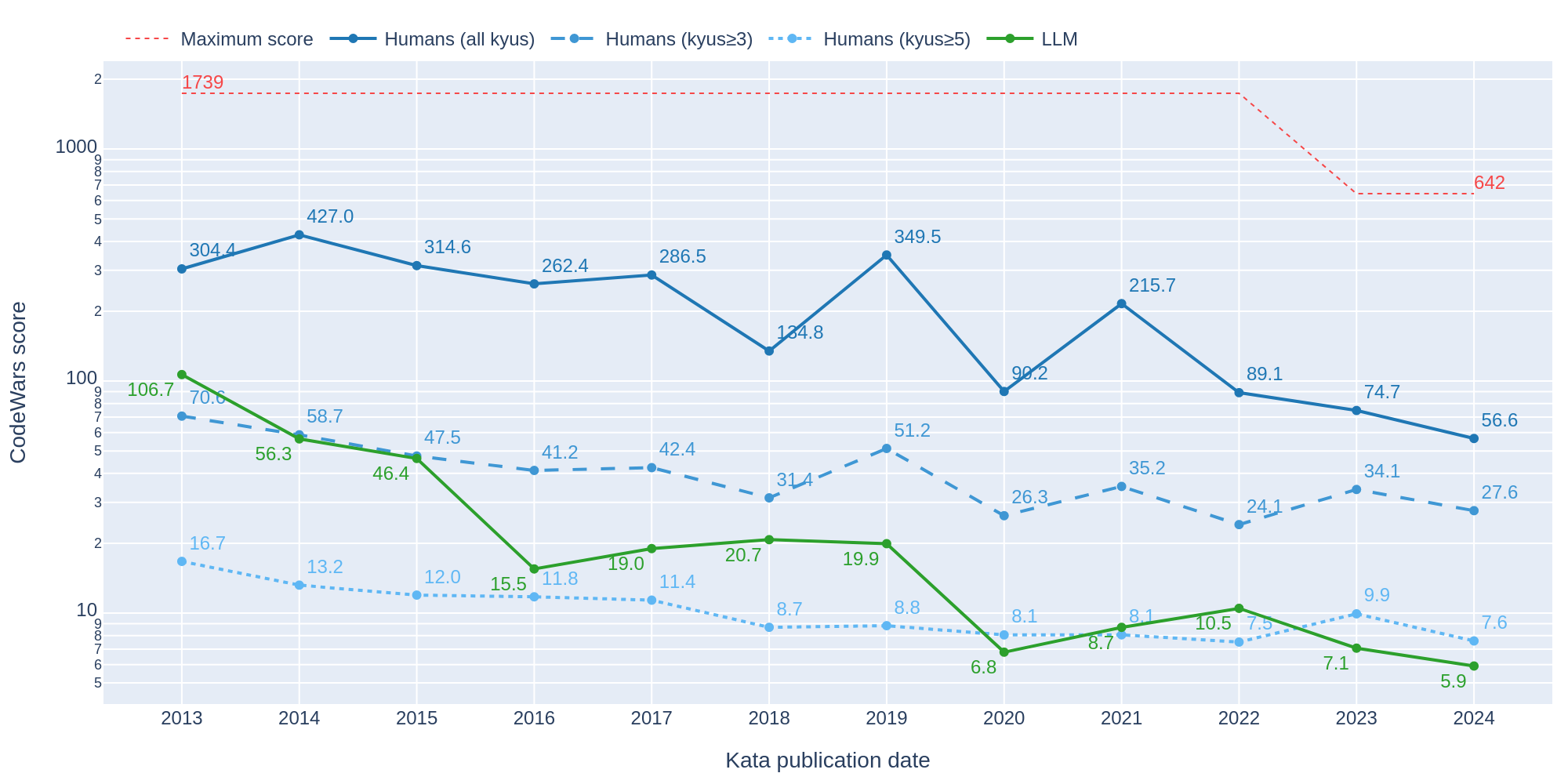}
    \caption{Codewars scores for the LLM and humans, on average, based on the publication date of the katas solved. Several trends for human performance are included, depending on whether the calculation considers katas of any level, or if high-complexity katas (kyu 1-2) or advanced and highly complex katas (kyus 1-4) are filtered out. The maximum score obtainable in each year is also shown, to account for no kyu 1 katas being published in 2023 and 2024.}
    \label{fig:timeScore}
\end{figure}

The variability in the LLM's capabilities is not limited to task difficulty or the language used. Surprisingly, the publication date of the kata also impacts performance, both for the LLM and for human users, as shown in Figure \ref{fig:timeScore}: \textbf{older katas are easier to solve than more recent ones}. This trend affects the score of human users, both globally and when excluding the most complex katas (focusing only on kyus $\ge$ 3), or when concentrating on basic and intermediate katas (kyus $\ge$ 5). However, this trend impacts the LLM even more significantly: initially, the LLM exhibits a higher score than humans if high-complexity exercises are ignored, but its performance eventually degrades to levels worse than those of human users in basic and intermediate exercises.

We consider two hypotheses that might explain this phenomenon:

\begin{enumerate}
    \item The criteria for selecting a kata's kyu level have changed, and a kata with a certain kyu $k$ in 2024 is more difficult to solve than a kata from 2013 with the same kyu $k$.
    \item Over time, solutions to katas become more widely available, allowing humans to copy those solutions, and LLMs to incorporate them into their training data.
\end{enumerate}

While the obtained results show no evidence toward one hypothesis or the other, a simple search for "codewars" in GitHub reveals approximately 38,500 public repositories containing kata solutions. This wide availability of solutions likely benefits both humans and LLMs, but it also suggests that newer katas, whose solutions have had less time to be widely spread, present a greater challenge for LLMs, as they lack copies of the solution in their training data. 

Previous works have already made similar observations about possible leakage of evaluations datasets through public repositories:  for instance, the SWE-bench dataset is actually built from public and widely known GitHub repositories, hence any recent LLM has probably seen part of the solutions \cite{swebenchverified}. Solutions for the programming challenges included in the APPS dataset have also been observed to be available in public repositories \cite{chen2021evaluatinglargelanguagemodels}. And more broadly, the Open LLM Leaderboard for general LLM performance has also required an update recently due to dataset contamination \cite{open-llm-leaderboard-v2}.

\subsection{High-level explanatory analysis}

Striving for a more quantitative explanation of the factors behind the LLM success at coding tasks, we present in this section a more detailed explanatory analysis.

Obtaining explanations from a neural network with billions of parameters such as the ones employed in LLMs is a complex task. The calculation and analysis of SHAP coefficients \cite{shap} is a popular approach to obtain explanations of the predictions of any kind of machine learning model in terms of its input features. However, the calculation of the SHAP coefficients requires computing how the absence of different groups of features affect the model output, something that can be achieved approximately by invoking the model a large number of times over different perturbations of the input features. Unfortunately this is very computationally demanding when dealing with an LLM, an also economically prohibitive in the case of a closed-source model like the one used in this paper. Furthermore, the SHAP coefficients provide explanations in terms of the input features, which in the case of the LLM are input tokens. As faithful as it might possibly be, an explanation of the LLM programming capabilities in terms of the presented tokens would be of little use.

What we propose here instead is to create a simple surrogate model that tries to predict the success of the LLM at solving a kata, making use of high-level features abstracting the nature of the kata and the programming language to be used, instead of the actual kata description. More precisely, the surrogate model is fed with the following features:

\begin{itemize}
    \item \textbf{Kata rank} (kyu).
    \item \textbf{Days since publication}: number of days elapsed since publication date, up to the date of the data collection (log-scaled).
    \item \textbf{Total Completions}: number of users who have completed the kata, regardless of the programming language used. (log-scaled).
    \item \textbf{Total Completion Rates}: fraction of users who have completed the kata among those who attempted it, regardless of the programming language used.
    \item \textbf{Same Language Completions}: number of users who have completed the kata in the same programming language that is being used in the attempt (log-scaled).
    \item \textbf{Same Language Completion Rates}: fraction of users who have completed the kata among those who attempted it in the same programming language that is being used in the attempt.
    \item \textbf{Language}: binary features marking the language that is being used in the attempt.
    \item \textbf{GitHub Language Pushes}: number of pushes in GitHub from 2020, for the language that is being used in the attempt (log-scaled).
    \item \textbf{GitHub Language Rank}: position of the language that is being used in the attempt in the GitHub ranking for most pushed programming languages.
\end{itemize}

Including features for both the ratio and the absolute number of users that completed a kata might seem redundant, but we hypothesize they could account for different underlying factors. On the one hand, the more users have solved the kata, the higher the probability of a solution getting leaked, and thus picked up during the LLM training. On the other hand, the ratio of users completing the kata might be a good estimate for the actual hardness of the kata within its kyu.

Two machine learning models were tested: a linear Support Vector Machine (SVM) \cite{liblinear} as implemented in scikit-learn \cite{scikitlearn}, and an XGBoost ensemble \cite{xgboost}. We performed a 5-fold cross validation to find the model hyperparameters combination resulting in the best accuracy, using the grid of parameter values show in Table \ref{tab:paramgrid}. Since an exhaustive grid search across all the hyperparameters for XGBoost would be time-consuming, a Randomized Search for 50 iterations was performed \cite{randomsearch}.

\begin{table}
    \centering
    \begin{tabular}{c|c|c}
        \textbf{Model} & \textbf{Parameter} & \textbf{Values} \\
        \hline
        \textbf{Linear SVM} & Penalty on training errors (C)& 20 uniformly spaced values in the log-range $[10^{-4}, 10^4]$ \\
        \hline
        \textbf{XGBoost} & Number of estimators & 1000\\
         & Penalty on number of leaf nodes ($\gamma$) & $[0, 10^{-5}, 10^{-4}, 10^{-3}, 10^{-2}, 10^{-1}, 1, 10, 100]$\\
         & Max tree depth & $\lbrace 6, 9, 12 \rbrace$\\
         & Data subsampling ratio & $\lbrace 0.5, 0.9, 1.0 \rbrace$\\
         & Column subsampling ratio (by tree) & $\lbrace 0.5, 0.9, 1.0\rbrace$\\
         & Penalty on large predicted values ($\lambda$) & $\lbrace 0, 1e-3, 1e-2, 1e-1, 1, 10, 100\rbrace$\\
    \end{tabular}
    \vspace{2mm}
    \caption{Grid of hyperparameter values used for optimizing the surrogate model for explainability.}
    \label{tab:paramgrid}
\end{table}

After hyperparameter optimization, the linear SVM model produced a cross-validation accuracy of $74.88\%$, while the XGBoost model attained a $74.91\%$. Being that the more complex ensemble model produces virtually the same results as the simpler one, we decided to discard the XGBoost model and perform the explanatory analysis over the linear SVM model.

A common issue that arises when analyzing linear models is multicollinearity. The data prepared for this study certainly includes correlations between features: the ratio of users that complete a kata is correlated with its kyu, and the number of users that complete a kata in a particular language is correlated with its rank on GitHub. Although a recommended approach to deal with collinearity is to remove correlated features, doing so in this case would obfuscate some possibly relevant effects already outlined: the rate of user completions might be a good estimate for the actual hardness of the kata within its kyu, and the number of users an estimate for the probability of solution leakage. Therefore, instead of pruning our feature set, we opt to make use of an alternative approach based on SHAP values.

\begin{figure}
    \centering
    \includegraphics[width=1\textwidth]{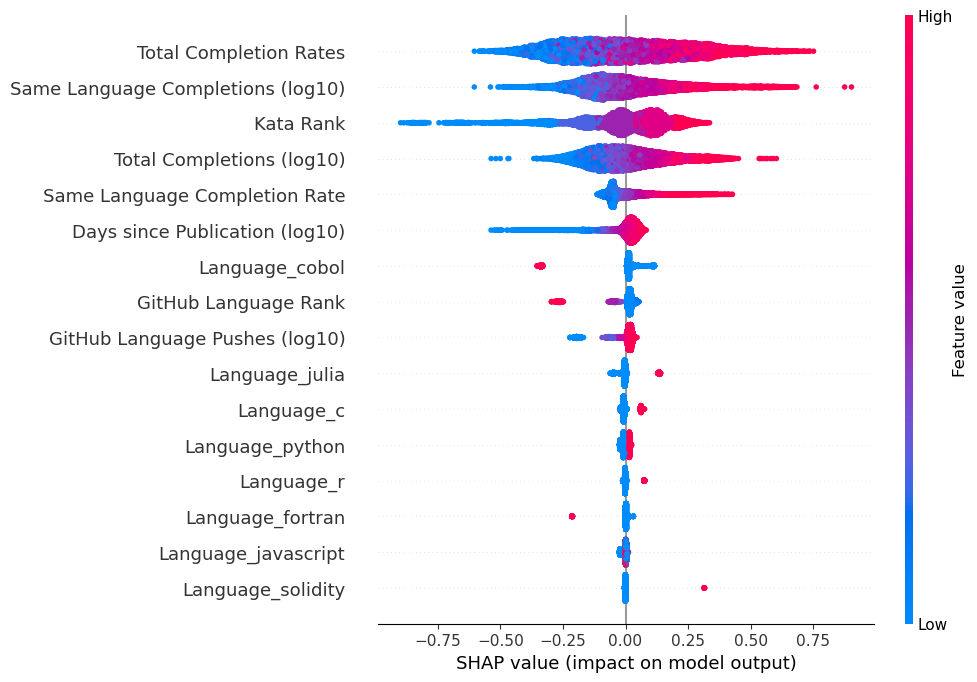}
    \caption{SHAP values of each explanatory variable in a surrogate model that attempts to predict the LLM's success in solving a kata. Each point represents an attempt to solve a kata in a specific language. Points to the right of the central axis indicate that the model assigns a higher probability of LLM success, while points to the left reduce that probability. The color of the point denotes the value of the explanatory variable being represented.}
    \label{fig:shap}
\end{figure}

Given a linear model in the form $f(x) = \sum_{i=1}^{d} x_i \beta_i + b$, the standard (or interventional) SHAP values for a sample $x$ and feature $j$ are simply given by $\beta_j (x_j - \mu_j)$, with $\mu_j$ the mean of feature $j$ across the data. These easy to compute estimates can only be obtained by assuming independent features, something that is not met in the case of multicollinearity. To address this, a non-interventional approach that does not assume feature independence can be used, where the SHAP coefficients are computed approximately via a sampling procedure, and a better credit assignment is given to correlated features \cite{shapLinear}. Through a sampling of 100K points we arrive at the SHAP coefficients represented in Figure \ref{fig:shap}.

The most relevant factors for the LLM's success include several variables that encode how many users have already completed the kata, as well as its difficulty (rank). Less impactful are the number of days since the kata was published, followed by all variables related to the language used. These results reinforce the descriptive analyses presented above.

\begin{figure}
    \centering
    \includegraphics[width=1\textwidth]{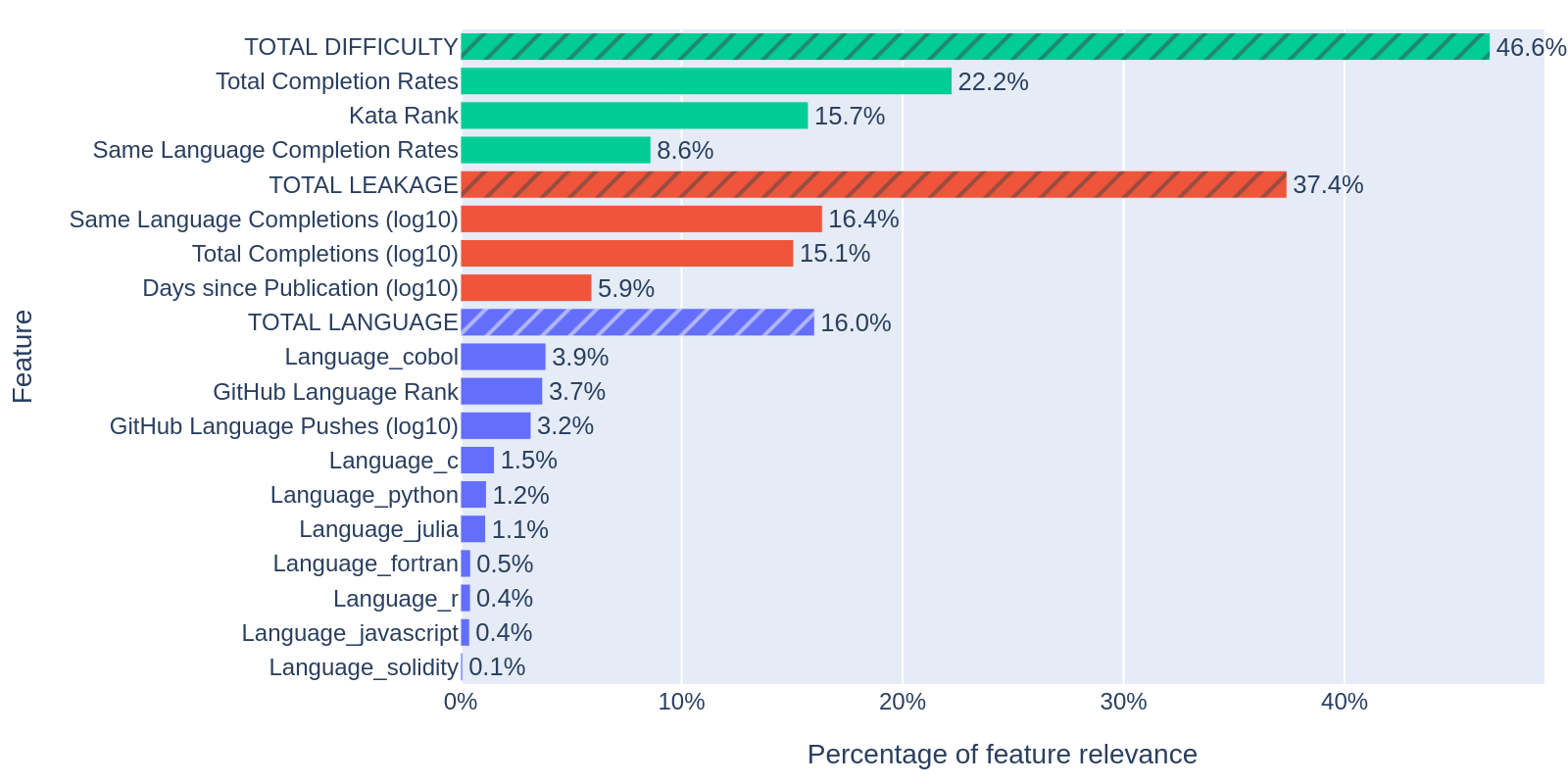}
    \caption{Aggregate of absolute SHAP coefficients for each surrogate model feature, which can be understood as the overall influence of each feature in the model, normalized to $100\%$. Features are grouped by the main underlying factor they encode (difficulty, leakage and language). The sums of coefficients within each group are also presented as striped bars.}
    \label{fig:shapAggregate}
\end{figure}

For a more quantitative analysis on the impact each feature has on the model, the absolute values of the obtained SHAP coefficients can be analyzed, as they account for how much each feature impacts each prediction, either increasing or decreasing its value. Summing up these values for each feature produces a metric for the total impact that feature has in the surrogate model predictions. Figure \ref{fig:shapAggregate} presents such values, normalized in the form of percentages.

Furthermore, we can cluster the features into three major groups:

\begin{itemize}
    \item Features that mainly encode the \textbf{difficulty} of the task: Kata Rank, Total Completion Rates and Same Language Completion Rates.
    \item Features that mainly encode the likelihood of a solution \textbf{leakage}: Days since Publication, Total Completions and Same Language Completions.
    \item Features that mainly encode the effect of the programming \textbf{language}: Language dummy features, GitHub Language Pushes and GitHub Language Rank.
\end{itemize}

Obviously none of this features encodes exclusively the factor of the group it was assigned to; e.g. Kata Rank mainly encodes difficulty, but katas with low Rank have fewer user completions, and thus a reduced likelihood of solution leakage. Nevertheless, we believe this clustering, even if simple, can provide some insight into the factors that drive the success of the LLM.

With this in mind, Figure \ref{fig:shapAggregate} also shows the total effect over the model of these groups of features. The \textbf{difficulty of the kata accounts for 46.6\%} of the model behavior, while \textbf{37.4\% is due to} features related with \textbf{solution leakage}, and the remaining \textbf{16\% depends on the programming language}.
It is not surprising to verify that the difficulty of the task accounts for a large part of the LLM performance. However, the fact that more than one third of the model behaviour is governed by features hinting a solution leakage is very revealing: the generalization capabilities of the LLM are limited by the amount of overlap between the training set and the tasks in which the LLM is being evaluated. 

It has been previously hinted \cite{hendrycks2021measuringcodingchallengecompetence} that the behavior of an LLM cannot be entirely attributed to memorization, as if that were the case the same success rate would be observed for easy and hard katas. However, we argue that easy katas can be solved by more humans, hence resulting in a higher probability of a solution leakage, even of many variants of the solutions. Other prior benchmarks have also pointed out that programming tasks involving popular Python libraries on GitHub seem to be easier to solve, although the sheer difficulty of using those libraries can overshadow the effect of their popularity \cite{wang-etal-2023-execution}; this also resonates with our results.


\section{Limitations}

In the light of the reviewed previous work on this topic and the general problem of LLMs for code generation, we must acknowledge the following limitations in this work.

\textbf{Coverage of actual software development tasks}: actual work as a software developer involves not just solving programming challenges, but also writing documentation, finding and reporting bugs in the system, designing high-level changes that span multiple code modules, gathering requirements and formalizing feature requests from users, organizing tasks with other team members, and defining the roadmap of the application. Thus, the presented evaluation only covers a small spectrum of this wide range of tasks. In addressing this shortcoming more comprehensive benchmarks such as SWE-bench \cite{jimenez2024swebenchlanguagemodelsresolve} have already been proposed.

\textbf{Reproducibility}: although the presented evaluation framework is fully automated, current terms of use of Codewars admonish the use of AI methods to gain ranks in their community. Therefore, we have decided to release neither the code of the developed botnet, nor the database of solutions proposed by the LLM, as publishing these artifacts would easen the misdeeds of those willing to incur in such fraudulent activities.

\textbf{Human in the loop}: previous work has noted that an LLM for code generation can improve its results if it takes part in a multi-turn conversation with a human that points out mistakes in the generated code \cite{austin2021programsynthesislargelanguage}. Commercial systems that make use of LLMs as helpers for coding tasks also make use of this loop \cite{gpt-4o, githubcopilot, gitlabduo}.

\textbf{Explainability}: while we have strived to implement an explainability analysis as faithful as possible, it is unrealistic to expect a simple model to encompass a complete representation of a closed-source LLM. The presented cross-validation analysis has revealed that our proxy model has an accuracy of $74.88\%$ when trying to predict the success of the LLM in solving the programming task, and thus the estimated influence levels of each factor contain some inherent noise due to proxy model error. In order to prevent further noise amplifications we have made use of a non-interventional approach to compute SHAP values. The development of more precise explainability techniques or the introduction of more meaningful high-level features of analysis might provide better estimates of the impact of difficulty, leakage and programming language on the LLM success.


\section{Conclusions and discussion}

In this paper we have reviewed the current methods for evaluating the capabilities of Large Language Models for code generation tasks, and carried out a new evaluation of GPT-4o-mini focused on programming challenges created by the Codewars community. Our analysis reveals several points of relevance:
\begin{itemize}
    \item \textbf{The LLM is unable to propose valid solutions to highly complex exercises}, and presents diminished performance when compared to humans when attempting hard problems.
    \item \textbf{Capabilities vary significantly across programming languages}, with major drops in performance for unpopular languages. Legacy languages are specially affected.
    \item \textbf{Leakage of solutions seems to account for up to 37.4\% of the performance}. Furthermore, a significant factor of leakage is the publication date of the challenge, with the LLM performing worse on more recent challenges.

\end{itemize}

Up to the best of our knowledge, this is the first result that quantifies the impact of solutions leakage on the performance of an LLM for coding, even if through approximate methods.

These results raise concerns about the current evaluation practices using public datasets, either in the form of crawls from public repositories or hand-made by crowdworkers.  Even if the research community continues to create larger or more curated evaluation datasets, these will be eventually integrated into the massive corpus used for LLM training; as a consequence newer models will deceptively perform better in the evaluations due to data contamination. In fact, current evaluations in the literature of the performance of state-of-the-art LLMs are, quite probably, overestimates of their real skill. While this phenomenon has already been pointed out in the LLM community for benchmarks like the Open LLM Leaderboard \cite{open-llm-leaderboard-v2}, in this paper we have wrought attention over the particular task of code generation and strived to quantify its impact.

As a proposal for improvement on evaluation methodologies, a fairer approach would be to evaluate new LLMs on datasets published strictly after the model training cut-off date. While we understand this requires significant efforts in the constant generation and tracking of new datasets, we also believe it would be feasible by generating new versions of existing evaluations datasets through iterative updates from their sources, e.g. incorporating new GitHub repositories in datasets like SWE-Bench, or new code challenges created by the community in the case of datasets such as the one presented here. These efforts would be of great value to improve the quality of current LLM evaluations, which in turn are paramount to identifying which aspects of LLM design and training constitute actual improvements in the field.

\section*{Acknowledgements}

We would like to give our thanks to the Codewars community for creating such an enriching environment for honing software craftsmanship skills, and in particular to the contributors Kayleigh and Hobovsky for useful discussions.

We also extend our thanks to our colleagues at IIC for providing comments on this work.

\printbibliography 

\end{document}